\documentclass[aps,prb,superscriptaddress,amsmath,floatfix,showpacs,twocolumn]{revtex4}
\usepackage{graphicx}  
\usepackage{subfigure}
\usepackage{amssymb}
\begin{document}

\title{Nanoscale ferromagnet-superconductor-ferromagnet switches controlled by magnetization orientation}
\author{Klaus Halterman}
\email{klaus.halterman@navy.mil}
\affiliation{Physics and Computational Sciences, Research and Engineering Sciences Department, Naval Air Warfare Center,
China Lake, California 93555}
\author{Oriol T. Valls}
\email{otvalls@umn.edu}
\affiliation{School of Physics and Astronomy and Minnesota Supercomputer
Institute, University of Minnesota, Minneapolis, Minnesota 55455}
\date{\today}



\begin{abstract}
We study clean ferromagnet-superconductor-ferromagnet (FSF) nanostructures
in which the magnetization of the  F layers can be parallel (P)
or antiparallel (AP). We consider the case where the thickness
of the S layer is of order of the coherence length, with  thinner
F layers. We  find that reversing the direction 
of the magnetization in one of the F layers leads in general to
drastic changes in the superconductor's state. Under a wide
variety of conditions, the AP geometry favors superconductivity. Magnetization
reversal in one of the F layers can lead to the superconductivity
turning on and off, or to  switching between different states.
Our results are obtained via self consistent solution
of the Bogoliubov-de Gennes equations and evaluation of the
condensation energies of the system. 
\end{abstract}

\maketitle



\paragraph*{Introduction:}
\label{intro}
Within the emerging field of spintronics\cite{igor}
considerable  interest has developed in devices in
which proximity effects are used to control
the superconductivity via  the
spin degree of freedom in ferromagnet (F)
and  superconductor (S) layered systems.
A large part of the motivation 
for this interest follows from earlier studies of systems
that involve nonmagnetic normal metal layers sandwiched between two 
ferromagnetic layers\cite{igor1} (FNF geometry).
In such devices 
the resistance of the system can change substantially
in the presence of a perturbing magnetic field. This
change  mainly arises from the 
spin-dependent scattering at the interfaces.
The ensuing giant magnetoresistance (GMR)  effect is found in 
spin-valves and magnetic multilayers 
where the relative orientations of the magnetization  in 
alternate ferromagnetic layers 
change as a function of an applied field.
If the local magnetization orientations are antiparallel (AP)  
the scattering will be stronger for a particular spin component, 
but if the magnetization vectors are aligned the more weakly 
scattered spin component 
carries the current 
with a lower resistivity.

If the 
nonmagnetic insert is 
replaced by a  thin superconductor, resulting
in a ferromagnet-superconductor-ferromagnet (FSF) junction,
a different type of spin-valve or spin-switch can be created\cite{gu}.
The
proximity effects arising from the mutual influence of the magnetic and
superconducting order parameters embody a variety of novel spin-valve effects
and device concepts, 
including high density nonvolatile memory,\cite{oh}  and magnetic
sensors.
The mechanism behind such devices is 
ultimately based\cite{tagirov} on the damped oscillatory nature of the
superconductor order parameter in the F regions, and the
associated magnetic correlations and destruction of superconductivity in the
S layer.  
In the transport regime, and with AP alignment
of the magnetizations in the F layers, a nonequilibrium spin
density can accumulate in the superconductor\cite{taka,zheng,yama}, destroying
the gap and resulting in a higher resistance state for a given
temperature.\cite{pena} Thus the superconducting correlations are controlled by
the relative magnetization orientation in the F layers.  Also,
quasiclassical thermodynamic considerations indicate that the transition
temperature, $T_c$, can be modified in a controlled way, thus allowing
supercurrent to flow in a predictable manner,\cite{tagirov,kulic,baladie,you}
yielding another type of spin switch.  
The superconducting order parameter and
$T_c$ in this case are again greatest when the the magnets are in the AP 
configuration, a
result shown to hold  for atomic thickness FSF layers as well.\cite{tollis2}
When the
superconducting system goes normal, the phase transition is second order for AP
magnetizations in the F layers and can be first order for parallel (P)
magnetizations if the  F layers are thin enough and the interface
transparency is high.\cite{tollis} If the outer ferromagnets are 
semiconducting
insulators, the $T_c$ variations have different signatures depending on whether
the superconductor is in the singlet or triplet state.\cite{kulic} 
An absolute
spin-valve effect can occur at spin-active interfaces in which the tunneling
current is finite for a range of voltages.\cite{hernando}

These types of devices are
in general most effective, and the effects most pronounced, for
junctions with clean interfaces and thin 
superconductors.\cite{baladie}
The lithographic, sputtering, and epitaxial methods
used in spin-switch fabrication permit
the creation of structures
as thin as a few atomic layers that have atomically flat interfaces. 
Moreover, high quality magnetic and nonmagnetic metallic films 
with an electron mean free path exceeding 150 \AA\ can also be readily 
fabricated.
One of the earliest experiments using FSF junctions involved CuNi/Nb/CuNi 
films,
and a magnetization direction dependence on $T_c$ was reported.\cite{gu}
A  spin switch was recently investigated using 
$\rm{La_{0.7}Ca_{0.3}MnO_3/YBa_2Cu_3O_7}$ superlattices 
that had large magnetoresistance
when in the superconducting state.\cite{pena} Spin valve
core structures involving  Nb/CuNi sandwiches
have also very recently\cite{pot} been reported. 
It is possible, in an FSF sandwich with AP magnetizations, 
for the electron in 
one of the magnets
to be Andreev reflected as a hole of the opposite spin in the other 
ferromagnet.\cite{melin}
This process of crossed Andreev reflection is believed to be behind
the results of 
experiments involving subgap transport
in Al/Fe hybrids.\cite{beckmann}
An  enhancement of the critical current and $T_c$ in Nb/Co was observed
and was attributed to a reduced exchange interaction in the domain structure of Co.\cite{kinsey}
Likewise, a type of spin-switch involving Nb/Permalloy layers 
revealed through transport measurements,
a decrease in the suppression of superconductivity.\cite{rusanov}
It was argued that the 
superconductivity is increased when
the magnetic domains are oriented differently, effectively
averaging in a way that reduces the effects of the exchange field.

Spurred by these important advances, we investigate 
here 
the effect of reversing one of the
magnetizations in clean FSF nanojunctions. We consider
the relevant case where  
the coupling between 
the S and F regions is appreciable, namely 
a thin S layer (of order of the BCS coherence length $\xi_0$) and relatively
large
magnetic exchange fields.
Our results are 
based upon numerical self consistent\cite{hv3} 
solution of the   microscopic Bogoliubov de-Gennes (BdG) 
equations. This method is most appropriate 
for the situation described above.
We calculate the pair
potential $\Delta({\bf r})$, the condensation energy, and the local density
of states (LDOS) for both the P and AP magnetization
configurations, over a range of values of the relevant parameters.
Our analysis will 
demonstrate   that under many conditions
the system can be made to switch from a superconducting state 
to a normal one, at low temperatures, by flipping the 
(collinear) magnetization orientation in one of the F layers, 
which can be achieved via
an applied field.
This will be illustrated by calculating the condensation energy 
as a function of ferromagnet thickness and exchange energy.
We find that the AP state is always the lowest energy state, and thus the
most favorable.
We conclude that  the  proximity effects that occur with P 
magnetization 
in successive F layers become substantially modified 
when adjacent F layers have AP magnetization alignment.
The pair amplitude and LDOS also 
display experimentally discernible characteristics
that depend on whether the magnets are in the P or AP configuration.

\paragraph*{Methods:} \label{meth}
The 
equations relevant to the
microscopic theory of inhomogeneous
superconductivity are the Bogoliubov-de Gennes (BdG) 
equations. \cite{bdg} 
We consider here an FSF structure
translationally 
invariant in the $x-y$ plane,
with interfaces normal to the $z$ 
direction. We assume parabolic
bands with bandwidths $E_F$ in the S layer
and $E_{F\pm}\equiv E_F \pm h_0$ in the magnet, where
$h_0$ is the Stoner exchange field. In the P geometry the
sign of $h_0$ is the same in both layers, while in the
AP geometry it is\cite{not} the opposite. The dimensionless
parameter $I \equiv h_0/E_F$ characterizes the magnetic
strength.
We include interface
scattering characterized by delta functions of
strength $H$ (dimensionless strength $H_B\equiv mH/k_F$).
We have written the BdG equations in this
geometry and with these assumptions in  previous\cite{hv3,hv1} work,
where we have also discussed\cite{hv4} the specific 
numerical methodology we use. 
It is therefore unnecessary to repeat  these derivations here.  
The exact quasiparticle energies
and amplitudes
are thus obtained  
by repeated iteration
of the BdG equations and the associated
self consistency condition for  $\Delta(z)$.

Once the energy spectra and pair potential are found, the condensation 
free energy, ${\cal F}$, (or, in the $T \rightarrow 0$ limit
the condensation energy) can be calculated. This is in
principle  straightforward, although  numerically
quite difficult. We use\cite{hv4} a particularly
convenient approach\cite{been,kos} which yields for the
condensation energy  $\Delta E_0$ the result:
\begin{equation}
{\Delta E_0} = \sum_{n'} \epsilon^0_{n'} - \sum_n \epsilon_n +
\int^d_0 dz \frac{|\Delta(z)|^2}{g},
\label{free2}
\end{equation}
where $g$ is the BCS coupling constant in S, 
$d$ the total sample width, 
$\epsilon_{n}$ and $\epsilon^0_{n'}$ are the free-particle energy 
spectra corresponding respectively  
to the superconducting and normal 
($\Delta(z)\equiv 0$) states, 
and the indices denote the appropriate\cite{hv4}
quantum numbers. 
The sums are performed over energies less than the usual cutoff $\omega_D$. 
Similarly, from the calculated self-consistent
eigenvalues and eigenfunctions one
can calculate the LDOS, $N(z,\varepsilon)$. This quantity is discussed
below.

\paragraph*{Results:} \label{res}
\begin{figure}
\centering
\includegraphics[width=3in]{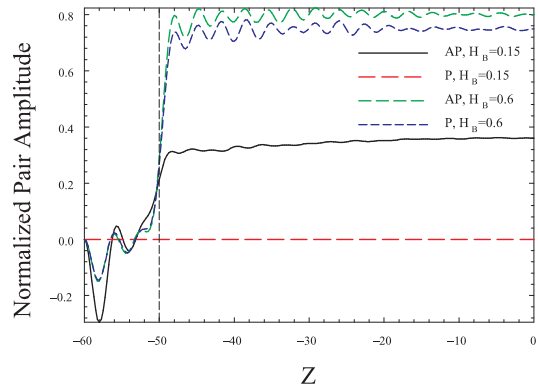}
\caption{(Color online) The spatially dependent  pair amplitude
(normalized
to its $T=0$ bulk value) for a 
$FSF$ trilayer, plotted vs. $Z \equiv k_{F}z$. The
contrasting  cases where the  magnetizations in the $F$
layers are parallel (label P), or antiparallel (label AP)
are shown. Because of symmetry, only
half of the sample is included. 
Each magnet has  width  $D_F \equiv k_F d_F =10$.
The dashed vertical line 
is at the F/S interface.
Two values of the scattering strength $H_B$ are considered. 
For the smaller value the pair amplitude vanishes
in the P configuration. 
The pair amplitude is always larger  
in the AP configuration. The exchange  parameter is $I=0.2$.
}
\label{fig1} 
\end{figure}

In the geometry we
consider, the inner superconductor layer of width $d_S$ is sandwiched between
two ferromagnet layers of equal width, $d_F$. 
The thickness $d_S$ is chosen
to be $d_S=\xi_0$, 
and we take $k_F \xi_0 = 100$ and $\omega_D=0.04 E_F$.
All results correspond to  low temperature, $T=0.01 T_c$.

\begin{figure}
\centering
\includegraphics[width=3in]{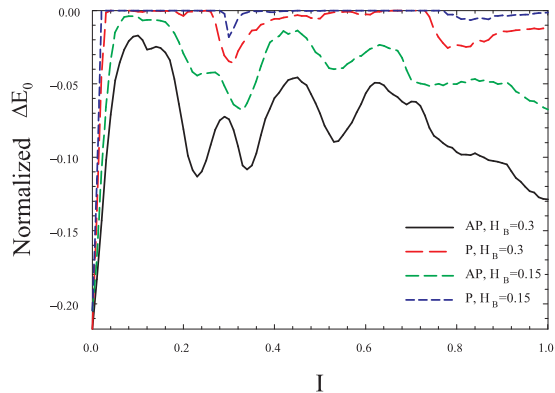}
\caption{(Color online) The condensation energy, $\Delta E_0$, 
normalized to $N(0)\Delta_0^2$, 
plotted vs 
$I$ for two values of
$H_B$.
The oscillatory behavior is complicated, but
for both  $H_B$ values the AP state is favored over the whole $I$ range. 
}
\label{fig2} 
\end{figure}

In Fig.~\ref{fig1}  we plot
the pair amplitude 
(the average $\langle \hat{\psi}_{\downarrow}({\bf r})
\hat{\psi}_{\uparrow}({\bf r})\rangle$, where the $\hat{\psi}_\sigma$ are the
usual annihilation operators), normalized to its zero temperature bulk value,
as a function of dimensionless distance.
Results are plotted both for both the P and AP
cases. 
Two values of the interface scattering 
parameter are considered, a small one ($H_B=0.15$) when
the proximity effect is strong, and a larger one
($H_B=0.6$) when it is weaker.
It is clear that the results
depend crucially on the relative magnetization
orientation, with the superconducting amplitude being
weakest in the P case. The effect is
magnified for the smaller $H_B$ value, 
where interface scattering is reduced.
In that case one can see in the figure that the
pair amplitude vanishes in the P case, while
being quite robust in the AP situation. 
The
superconductivity can thus {\it be switched on and off} by
reversing the magnetization in one of the F layers.

This
favoring of the AP configuration is, qualitatively,
a very general phenomenon. To see this,
it is very convenient 
to analyze the pair condensation 
energy (Eq.~(\ref{free2})).  
The trends in the pair amplitude, such as those
in Fig.~\ref{fig1}, should  be reflected
in the condensation energy,
which
should then be lower (higher in absolute value) in the AP
case than in the P configuration.
$\Delta E_0$ is shown in the next two figures,
normalized to $N(0)\Delta_0^2$ which is twice its 
bulk zero temperature value. 
In Fig.~\ref{fig2} this normalized quantity
is plotted as a function of $I$ for two values of $H_B$.
The F width is kept fixed at $D_F\equiv k_F d_F=10$
(recall that $D_S \equiv
k_F \xi_0$ always). The 
entire range of $I$ from the nonmagnetic
($I=0$) limit to the half metallic case ($I=1$) is spanned. 
For all nonzero $I$, the AP case is always favored.
This trend persists even for  larger $H_B$ (not shown)
where the proximity effect is weaker.
At $I=1$ only one spin band is populated at the Fermi level
and consequently $|\Delta E_0|$ is large, as Andreev reflection 
is depressed and the Cooper pairs are more restricted to the S region.
We also see in the figure that the {\it difference}
in condensation energies for P and AP geometries at fixed
thickness is a weak function of $I$, except  at small $I$.

\begin{figure}
\centering
\includegraphics[width=3in]{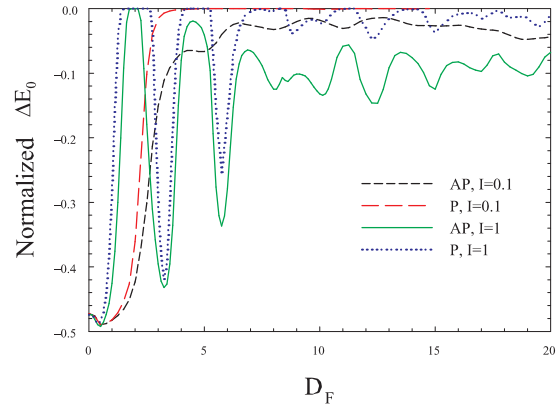}
\caption{(Color online) The normalized condensation energy vs. 
the dimensionless ferromagnet width
$D_F$. Two values of $I$ are considered, as indicated, at $H_B=0.3$. 
For a given $I$, 
the AP state has 
lower $\Delta E_0$ over the whole $D_F$ range included. 
}
\label{fig3} 
\end{figure}

The AP configuration continues to
be preferred when the thickness $D_F$ is varied at constant
$I$. This is shown in Fig.~\ref{fig3}, where we plot
the normalized $\Delta E_0$ versus $D_F$. Results for two values
of $I$ are plotted, and both P and AP configurations
are studied. 
As $D_F \rightarrow 0$, 
one is left only with the superconductor and $\Delta E_0$
approaches its bulk value. Increasing $D_F$
causes initially a sharp
rise in $\Delta E_0$.
The  condensation energy then  saturates, exhibiting
damped irregular oscillations,
reflecting the competition
between magnetism and superconductivity.  Again, in  all cases superconductivity favors the AP
configuration. At small $I$ ($I=0.1$),
$\Delta E_0$ for the P configuration  vanishes beyond $D_F\gtrsim4$,
while in the half-metallic limit, $\Delta E_0$ 
is an oscillatory function of $D_F$. The
difference in condensation energies between P and AP configurations
at the same $I$ is a weak function of $D_F$.

\begin{figure}
\centering
\includegraphics[width=3in]{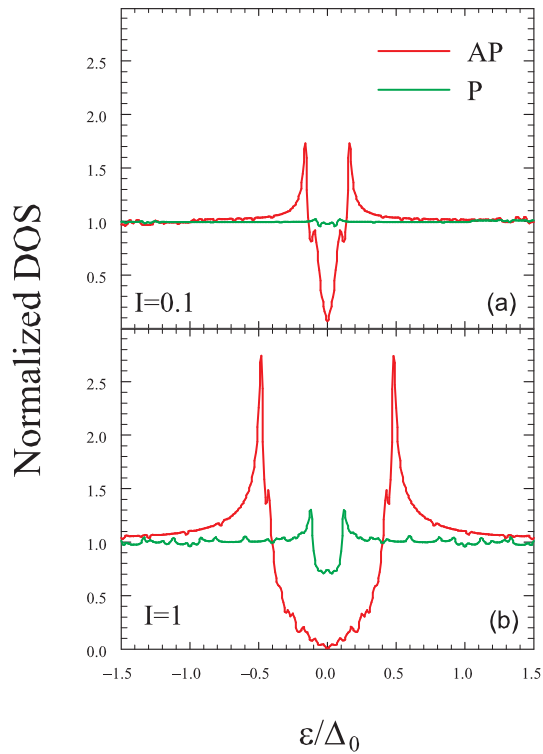}
\caption{(Color online) The normalized LDOS (at
$D_F=10$ and $H_B=0.15$)
spatially averaged over the S region and normalized
as explained in the text.
Results for both  P  and AP configurations
are shown for two $I$ values. 
}
\label{fig4} 
\end{figure}

The irregular oscillatory behavior 
in Figs.~\ref{fig2}-\ref{fig3}
reflects
the existence of the characteristic length 
$\ell=(k_\uparrow - k_\downarrow)^{-1}$
arising from the difference between  Fermi wavectors for up and down 
spins in the F layers. Such oscillatory behavior depends on
the relation between $d_F$ and $\ell$.
The
latter in turn
depends on $I$. Since 
the ratio $\ell / d_F$ depends on $D_F$ in a simpler way than on $I$, 
the oscillations
are best studied in Fig.~\ref{fig3}. There one can see that at larger 
$I=1$ (hence smaller $\ell$), 
the characteristic oscillations have a shorter spatial period than those at $I=0.1$. 
For $SFS$ sandwiches with small $D_F$
one finds\cite{hv1}  oscillations in the pair
amplitude in F. These can be seen in the
left edge of Fig.~\ref{fig1}. The situation for $\Delta E_0$ 
is  much more complicated, since oscillations
in the pair amplitude are only indirectly reflected there. 

The
strong modifications to the superconducting state
of the sample should be easily detected in measurements
of the critical current. These switching effects are also
very well reflected  in LDOS results. Thus, we show in Fig.~\ref{fig4}
the LDOS $N(z,\varepsilon)$ 
averaged over the S region. The results are normalized to
the normal state bulk value. We display results  for
two values of $I$, $D_F=10$, and $H_B=0.15$.
One can plainly see the difference
between P and AP configurations: in the AP case the BCS like peaks are
much more prominent and the gap fairly well defined.  In the P case no
gap exists, although a weak BCS like feature is still visible for $I=1.0$,
while the
features flatten out nearly completely at small $I$ ($I=0.1$), when 
the system is no longer superconducting, as
seen by the the vanishing of the condensation energy in this case, 
(Fig.~\ref{fig2}).

The enhancement of superconductivity in the AP configuration is
not, as one might naively think, simply a consequence
of the magnetic polarizations canceling in the
superconductor. As has been found,\cite{hv3} 
the magnetic moment induced in the superconductor
by the ferromagnetic contacts penetrates into the S material
only a few Fermi wavelengths. We have verified that this is also
the case here. Thus, the reasons are more subtle.
The weakening of superconductivity by ferromagnetic
contacts depends in a complicated way on the amplitudes
for electron scattering (both normal and Andreev)  
at the
interfaces. These are to a greater or lesser strength
pair breaking. The penetration depth
for Cooper pairs
into the F material is much smaller than for a normal metal
and this is reflected in the interface scattering amplitudes.
Our self consistent calculation shows then, that the
superconducting state (with $d_S=\xi_0$) can better 
survive proximity to two F contacts that have opposite
polarizations.



{\it Acknowledgments:}
 We thank Paul Barsic for several conversations. 
This work was supported in part by a grant of HPC resources 
from ARSC at the University of 
Alaska Fairbanks as part of the DoD High Performance 
Computing Modernization Program.


\begin{thebibliography}{0}
\bibitem{not} We consider only the collinear case for $h_0$, and no
spin-orbit coupling, so that only singlet pairing needs to be included.
\end{thebibliography}
\end{document}